# On the top of the energy barrier [*]

Margarita García Pérez [a], Antonio González-Arroyo [b], Jeroen Snippe[a] and Pierre van Baal[a]

[a] Instituut-Lorentz for Theoretical Physics,
University of Leiden, PO Box 9506,
NL-2300 RA Leiden, The Netherlands.

[b] Departamento de Física Teórica C-XI,
Universidad Autónoma de Madrid,
28049 Madrid, Spain.

We present results related to the search of SU(2) instantons on a geometry $[0, L]^3 \times [0, T]$ obtained using over-improved cooling [1] with fixed boundary conditions. We also introduce a criterion for finding a sphaleron at the top of the energy barrier of the instanton path.

## 1. INTRODUCTION

In a previous publication [1] we have developed a method to find stable instantons on a lattice. It is a well known fact [2] that the Wilson action does not allow for such configurations. The reason is that the lattice artefacts drive them to localized objects of the size of the lattice spacing which finally "drop" through the lattice. To cure the problem one can play with these lattice artefacts and define an action that has still the correct continuum limit and allows for stable instantons. The one we have selected, and used up to now in our numerical simulations, is

$$S(\varepsilon) = \frac{4-\varepsilon}{3} \sum_{x,\mu,\nu} \mathrm{Tr}\left(1 - \overset{\nu\ \Box}{\underset{x\ \mu}{}}\right) \qquad (1)$$
$$+ \frac{\varepsilon-1}{48} \sum_{x,\mu,\nu} \mathrm{Tr}\left(1 - \overset{\nu\ \Box\Box}{\underset{x\ \mu}{}}\right) \quad .$$

One can check (for details see [1]) that for a smooth continuum configuration ($A_\mu(x)$) defined on the lattice, the modified action of eq. (1) provides the correct continuum limit while leaving freedom to tune the effect of the lattice artefacts

---
[*]Based on the talks presented by the first and fourth author at Lattice'93 (Dallas, 12-16 Oct, 1993).

by tuning $\varepsilon$, i.e. ($a$ is the lattice spacing)

$$S(\varepsilon) = \sum_{x,\mu,\nu} \mathrm{Tr}\ \left[\ -\frac{a^4}{2}F_{\mu\nu}^2(x) + \frac{\varepsilon a^6}{12}(\mathcal{D}_\mu F_{\mu\nu}(x))^2\right.$$
$$\left. +\mathcal{O}(a^8)\right] \quad . \qquad (2)$$

Just by dimensional arguments one deduces for the continuum instanton

$$S(\varepsilon) = 8\pi^2(1 - \varepsilon\, d\ (a/\rho)^2 + \mathcal{O}(a/\rho)^4) \qquad (3)$$

where $\rho$ is the scale parameter of the instanton and eq. (2) implies $d > 0$. The Wilson action corresponds to $\varepsilon = 1$ and clearly gives rise to an action that decreases as $\rho$ decreases, explaining the instability. If, on the contrary, we choose $\varepsilon$ negative (which we will call over-improvement [1]) the instanton will grow to make the action decrease until it reaches the <u>maximum</u> size allowed by the finite (lattice) volume. This turns out to be extremely convenient, as we are interested in finding the largest instanton that fits in a volume $[0, L]^3 \times [0, T]$. This instanton will correspond, as $T \to \infty$, to tunnelling between two classical vacua (through presumably the lowest energy barrier) and we conjecture that the configuration that sits at the top of the instanton energy barrier is a sphaleron. The hope is, that knowing the sphaleron will allow us to map out the relevant degrees of freedom to compute the non-perturbative contributions to the spectrum (for further details and a similar study in $S^3 \times \mathbb{R}$ see [3] and references therein).

The search for this widest instanton is complicated by the fact that $\rho$ is not the only relevant parameter for the configurations. For the infinite volume instanton, for instance, there are in addition the 4 position and 3 global gauge parameters. The situation is rather different on a hypertorus $[0, L]^3 \times [0, T]$. It is known [4] that, as long as $T$ is kept finite, there are no regular self-dual solutions with topological charge 1 and periodic boundary conditions. The proof no longer holds when one introduces twisted boundary conditions [5] keeping at least one component of the twist tensor $n_{\mu\nu} \in \mathbb{Z}_2$, different from zero. Indeed, for orthogonal twist $\frac{1}{8}\epsilon_{\mu\nu\lambda\sigma}n_{\mu\nu}n_{\lambda\sigma} = 0 \mod 2$ and $n_{\mu\nu} \neq 0 \mod 2$, for some $\mu$, $\nu$, it has been shown [6] that solutions exist by gluing a localized instanton (with its 8 parameters) to the zero-action configuration in the presence of a twist. However, the proof only refers to small instantons and does not give much of a clue about the moduli space of large instantons that extend up to the "boundary" of the torus. In [1] we have presented results with both twisted and untwisted boundary conditions that shed some light on this question. Let us briefly summarize these results.

When $T \to \infty$ the action can only remain finite if the energy density vanishes at $|t| \to \infty$. From the vanishing of the magnetic energy one deduces

$$A_i(\vec{x}, t \to \pm\infty) = iC_i^{\pm}\sigma_3/2L \quad , \qquad (4)$$

where $C_i^{\pm} \in [0, 4\pi]$ ($A_0 = 0$). Moreover, the vanishing of the electric energy implies that $C_i^{\pm}$ has to be asymptotically time independent. We can therefore characterize instantons in $T^3 \times \mathbb{R}$ by the values of $C_i^{\pm}$. Assume now that the instanton configuration prefers to interpolate between non-equal points in the vacuum valley, that is $C_i^{+} \neq C_i^{-}$. If we make the time finite and impose periodic boundary conditions, the configuration will have to move along the vacuum valley to interpolate between $C_i^{-}$ and $C_i^{+}$. This movement will give rise to a non-zero and constant electric energy $\mathcal{E}_E(t)$, where the magnetic energy $\mathcal{E}_B(t)$ goes to zero, thus destroying the self-duality. As pointed out in [1] we have clearly observed this behaviour in the numerical simulations with periodic boundary conditions. For increasing $T$ the electric energy in the tail will decay as $1/(T - T_0)^2$ ($T_0$ being the time interval where $\mathcal{E}_B(t) \neq 0$) such that as $T \to \infty$ self-duality is restored. We have also checked that a twist in time removes the electric tail even at finite $T$, giving rise to 8-parameter solutions ($\rho$, position and the 3 parameters in the vacuum valley $C_i^{+} = 2\pi - C_i^{-} \pmod{4\pi}$ (up to a periodic gauge transformation)).

Several questions remained, such as the existence of instanton solutions for arbitrary values of $C_i^{\pm}$ and the interplay between $\rho$ and these boundary values.

In what follows we will present some preliminary results obtained using fixed boundary conditions (f.b.c.), a trick that will allow us to explore the vacuum dependence of the instantons. To conclude we will also introduce a criterion for finding a sphaleron at the top of the energy barrier along the instanton path.

## 2. FIXED BOUNDARY CONDITIONS

The gauge invariant observables that characterize a configuration in the vacuum valley are the Polyakov line expectation values

$$P_i \equiv \tfrac{1}{2}\mathrm{Tr}\left(\mathrm{Pexp}(\int_0^L A_i(\vec{x}, t)dx_i)\right) = \cos\left(\frac{C_i}{2}\right), \quad (5)$$

as well as the products

$$\tfrac{1}{2}\mathrm{Tr}\left(\mathrm{Pexp}(\int_0^L A_i(\vec{x}, t)dx_i)\mathrm{Pexp}(\int_0^L A_j(\vec{x}, t)dx_j)\right)$$
$$= \cos\left(\frac{C_i + C_j}{2}\right) \equiv P_{ij}, \quad i \neq j, \qquad (6)$$

We want to address the question of whether instantons exist interpolating between any two sets $\{P_i\ (t = \pm\infty) \equiv P_i^{\pm},\ P_{ij}\ (t = \pm\infty) \equiv P_{ij}^{\pm}\}$. For that we propose to perform over-improved cooling with the constraint of keeping the observables $\{P_i^{\pm}, P_{ij}^{\pm}\}$ fixed at the time boundaries of the lattice. This idea can be easily implemented on a $N_s^3 \times (N_t + 4)$ lattice by fixing the values of the link variables $U_i(\vec{x}, n_t = 1, 2) \equiv U_i^{-}$

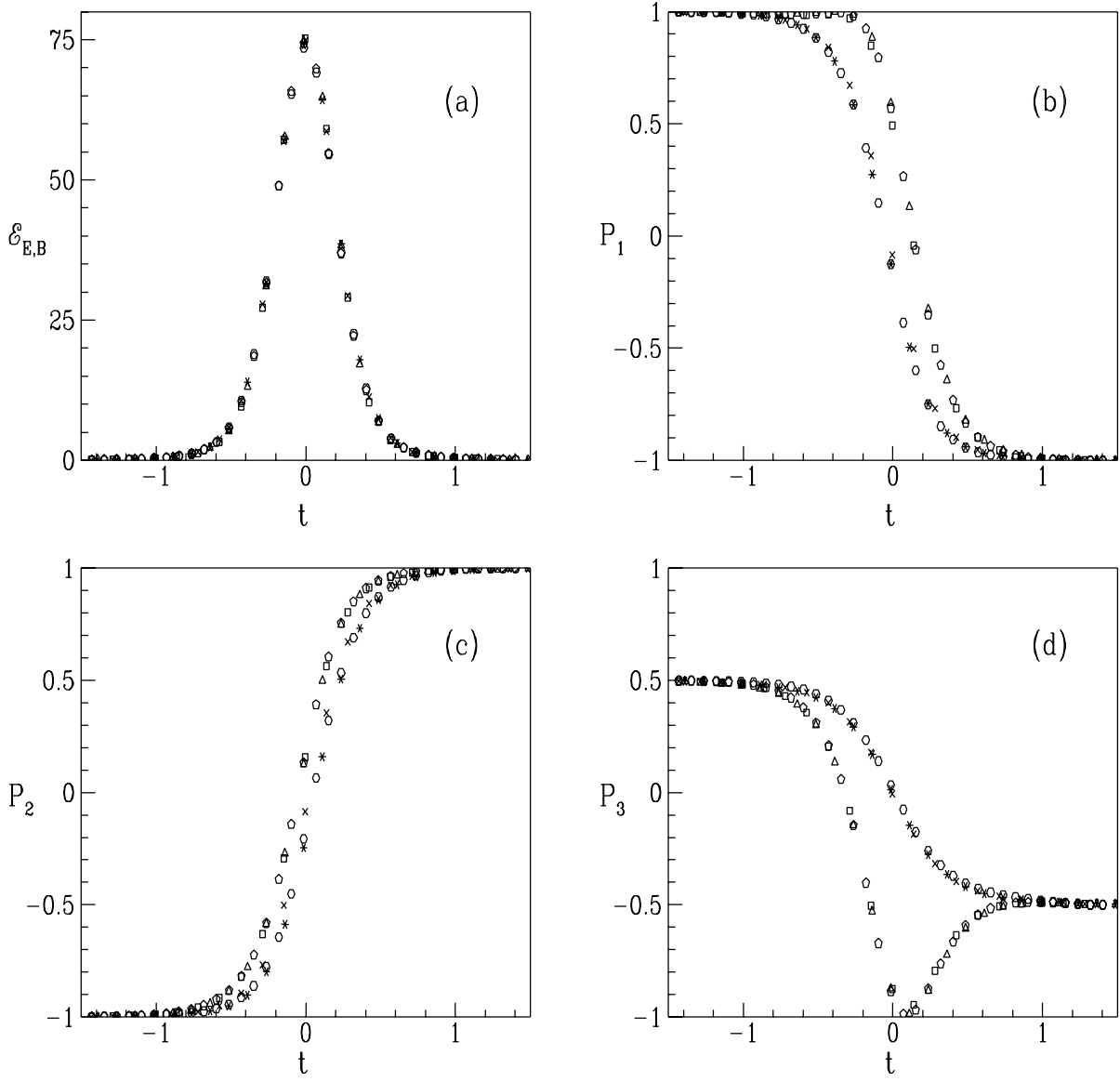

Figure 1. Numerical results for lattices $7^3 \times 21$ (squares), $8^3 \times 24$ (triangles) and $12^3 \times 36$ (pentagons) obtained from over-improved cooling with $\varepsilon = -1$ and fixed boundary conditions given by eq. (9). Fig. (a) contains data for $\mathcal{E}_E(t)$ with the above mentioned symbols and $\mathcal{E}_B(t)$ (crosses for $N_s = 7$, stars for $N_s = 8$ and hexagons for $N_s = 12$). In fig. (b-d) we plot the Polyakov lines $P_i(t)$ through the points with maximal (squares, triangles and pentagons) and minimal (crosses, stars and hexagons) $E_3^2$ at $t = 0$.

and $U_i(\vec{x}, n_t = N_t + 3, N_t + 4) \equiv U_i^+$, to be

$$U_i^\pm = \begin{cases} I & \text{if } n_i \neq N_s \\ \cos(\frac{C_i^\pm}{2}) + i\sigma_3 \sin(\frac{C_i^\pm}{2}) & \text{if } n_i = N_s \end{cases} \quad (7)$$

It is easy to check that this choice gives the correct result for $\{P_i^\pm, P_{ij}^\pm\}$. Notice also that it is necessary to fix the boundary configurations in 4 instead of 2 time slices, since $S(\varepsilon)$ (eq. (1)) contains terms proportional to $2 \times 2$ plaquettes that would induce a coupling between the two boundaries if we would update the links at $n_t = 2$ or $N_t + 3$. We also fix

$$U_0(\vec{x}, n_t = 1, N_t + 3) = I \quad . \quad (8)$$

This procedure therefore yields a $N_s^3 \times N_t$ lattice with fixed values of the time boundary conditions.

To show the viability of the method we present in fig. 1 the results for $\varepsilon = -1$ and

$$C^- = (0, 2\pi, 2\pi/3), \ C^+ = (2\pi, 0, 4\pi/3) \quad , \quad (9)$$

for lattices with $N_s = 7, 8, 12$ and $N_t = 21, 24, 36$ respectively. We plot (a) both the electric and magnetic energy ($\mathcal{E}_{E,B}(t)$) after appropriate scaling with $N_s$, and (b,c,d) the Polyakov lines $P_i(t)$ trough two particular points $\vec{x}$, corresponding to respectively maximal and minimal $E_3^2$ at t=0. The scaling properties of the configuration, as well as the good degree of self-duality of the solution are clear. It is useful to point out here that the instanton approaches the vacuum valley only exponentially, so that by forcing it to decay to a pure gauge for finite $T$ one distorts the configuration. However, choosing $N_t$ big enough with respect to $N_s$ makes the degree of distortion away from self-duality quite small, as is obvious from fig. 1(a).

In [1] we have noted that the numerical simulations with the twist $n_{0i} = (1, 1, 1)$ seemed to indicate that the largest instanton preferred to interpolate between the values of eq. (9). To check this further we have performed simulations on a $8^3 \times 24$ lattice, keeping $C_{1,2}^- = 0, 2\pi$ fixed, while changing $\cos(C_3^-/2)$ over the entire range $[-1, 1]$, with the twisting condition $C_i^+ = 2\pi - C_i^-, \forall i$. The initial configuration, with $\cos(C_3^\pm/2) = \mp 1$, was generated from a random start (with f.b.c.) to which we applied over-improved cooling with $\varepsilon = -1$ until the action was close to $8\pi^2$. From there we went on by changing in small steps the values of $C_3^\pm$, each step cooling again the configuration for 500 sweeps to eliminate the distortion due to the change of boundary values. In that way we generated a row of configurations with different values of the Polyakov lines $P_3^\pm$. If necessary each configuration was further cooled to improve the degree of self-duality. The corresponding results are presented in fig. 2. In 2(a) we plot $\mathcal{E}_{E,B}(t)$ and in (b,c,d) the Polyakov lines, this time just through the point with maximal $E_3^2$ at t=0. Fig. 2(a) shows how the central part of the energy density is just slightly modified while changing $C_3^\pm$, indicating that the dependence of the size of the instanton on certain choices of the vacuum parameters can be rather weak. This could be an indication that the lowest positive eigenvalue of the energy functional at the sphaleron is small. We have also blown up the tails in fig. 2(a) to show how certain choices of boundary conditions give rise to non-selfdual configurations, due to the appearance of an electric tail (which does not decrease under further cooling). However, as we previously indicated, as $T \to \infty$ this tail effect will disappear giving rise presumably to a self-dual instanton on $T^3 \times \mathbb{R}$.

## 3. SPHALERON CRITERION

A sphaleron is a solution of the 3-dimensional equations of motion with precisely one unstable mode, presumably along the tunneling path between the classical vacua. This means that the sphaleron satisfies in the continuum the equation

$$\sum_i \mathcal{D}_i F_{ij} = 0 \quad . \quad (10)$$

This equation can be easily rewritten on the lattice. Take for instance the 3-dimensional $SU(2)$ Wilson action

$$S_{\rm W} = \sum_{\vec{x}, i, j} {\rm Tr}\left(\mathbf{1} - \raisebox{-2pt}{\includegraphics[height=10pt]{plaq}}\right) \quad , \quad (11)$$

and replace the link $U_i(\vec{x})$ by $e^X U_i(\vec{x})$, with $X$ a Lie-algebra element. The condition of extremum

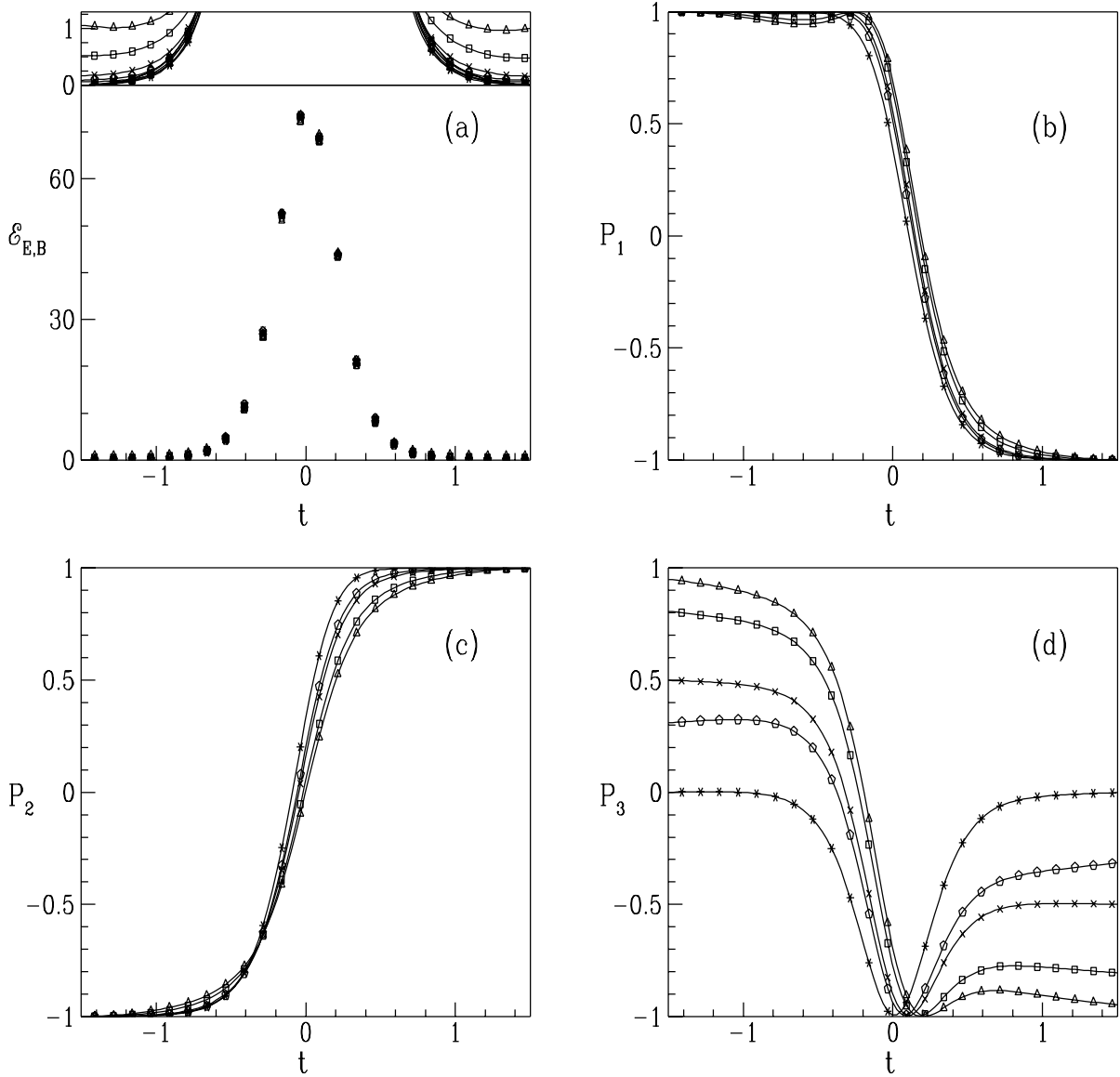

Figure 2. Numerical results for a lattice $8^3 \times 24$ obtained from over-improved cooling with $\varepsilon = -1$ and changing boundary conditions as described in section 3. Fig. (a) contains data for $\mathcal{E}_{E,B}(t)$, in the upper part of the figure we plot the tails at an enlarged scale. In fig. (b-d) we plot the Polyakov lines $P_i(t)$ through the points with maximal $E_3^2$ at $t = 0$.

of the action implies

$$U_i(\vec{x}) = \frac{\tilde{U}_i(\vec{x})}{\|\tilde{U}_i(\vec{x})\|} \quad , \qquad (12)$$

with

$$\tilde{U}_i(\vec{x}) = \sum_{j \neq i} \left( \begin{array}{c}\end{array} + \begin{array}{c}\end{array} \right) \equiv a_0 + i\vec{a}\cdot\vec{\sigma} \quad , \qquad (13)$$

and $\|\tilde{U}_i(\vec{x})\| = \sqrt{a_\mu^2}$. For the sphaleron configuration, eq. (12) has to be fulfilled. Let us introduce the following functional [7]:

$$\tilde{S} = \sum_{\vec{x},i,j} \mathrm{Tr}\left\{ \mathbf{1} - U_i(\vec{x})\frac{\tilde{U}_i^\dagger(\vec{x})}{\|\tilde{U}_i(\vec{x})\|} \right\} \quad . \qquad (14)$$

For a sphaleron, as well as for a pure gauge configuration, $\tilde{S} = 0$. The reasoning applies exactly the same for the over-improved action $S(\varepsilon)$ (eq. (1)) but in that case $\tilde{U}_i(\vec{x})$ also contains staples associated to the $2 \times 2$ plaquettes attached to the link $U_i(\vec{x})$. We will denote the corresponding functional by $\tilde{S}(\varepsilon)$. It will give us a perfect criterion to check whether or not the configuration that sits at the top of the instanton tunneling path is a sphaleron.

In fig. 3 we plot $\tilde{S}(\varepsilon = -1)$ (after appropriate scaling with $N_s^3$) for the instanton path of fig. 1. The shape changes dramatically at $t = 0$, seemingly indicating that we are close to a zero value for $\tilde{S}(\varepsilon = -1)$. However, for this particular choice of boundary conditions there is still a small non-zero gap at $t = 0$ that does not decrease while increasing $N_s$ (decreasing the lattice spacing $a$) and is thus not expected to be a lattice artefact. By monitoring the dependence of this gap as function of $N_t/N_s$ we have also verified that it is not due to the finiteness of $T$. All this indicates that the chosen values of the boundary conditions in fig. 1 are not yet exactly the ones that correspond to tunneling over the lowest energy barrier and that exploring further the vacuum dependence of the instanton configurations is necessary. We also intend to use minimization of $\tilde{S}(\varepsilon)$ [7] to directly find the true sphaleron.

## ACKNOWLEDGEMENTS

This work was supported in part by grant number AEN 90-0272, financed by CICYT, and by grants from "Stichting voor Fundamenteel Onderzoek der Materie (FOM)" and "Stichting Nationale Computer Faciliteiten (NCF)" for use of the CRAY Y-MP at SARA. M.G.P. was supported by a Human Capital and Mobility EC fellowship.

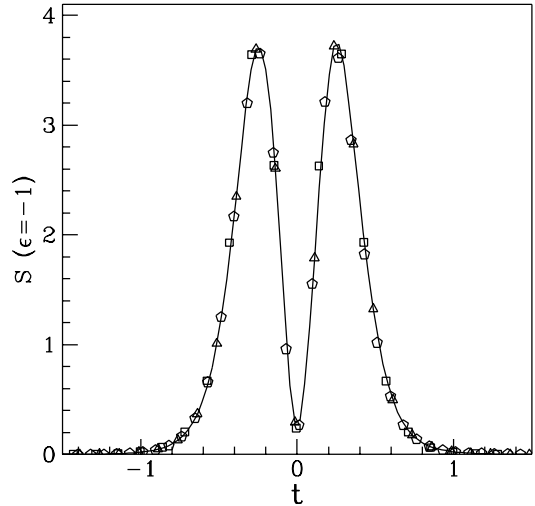

Figure 3. We plot $\tilde{S}(\varepsilon = -1)$ (after scaling with $N_s^3$) along the instanton tunneling path of fig. 1. Data correspond to lattices $7^3 \times 21$ (squares), $8^3 \times 24$ (triangles) and $12^3 \times 36$ (pentagons).